\documentclass[twocolumn,10pt]{article}
\usepackage[margin=1.9cm,top=2.1cm,bottom=2.1cm]{geometry}
\usepackage{graphicx,booktabs,amsmath,amssymb,microtype,xcolor}
\usepackage[hidelinks]{hyperref}
\usepackage{caption}
\usepackage{titlesec}
\titlespacing*{\section}{0pt}{9pt}{4pt}
\titlespacing*{\subsection}{0pt}{6pt}{2pt}
\titleformat{\section}{\normalfont\bfseries}{}{0em}{}
\titleformat{\subsection}{\normalfont\itshape}{}{0em}{}
\title{\vspace{-8mm}\Large\textbf{Copying explains the collective behavior of AI agents in the wild}}
\author{\normalsize Giordano De Marzo$^{1,2,\ast}$, Nicola Albor\'e$^{3,\ast}$ and David Garcia$^{1,4}$ \\[3pt]
\small $^1$University of Konstanz, Konstanz, Germany \quad $^2$Centro Ricerche Enrico Fermi, Rome, Italy \\[1pt]
\small $^3$Intesa Sanpaolo, Data \& Artificial Intelligence Office, Turin, Italy \quad $^4$Complexity Science Hub, Vienna, Austria \\[2pt]
\small $^\ast$These authors contributed equally to this work.}
\date{}
\begin{document}
\maketitle
\begin{abstract}\noindent
In June 2026, thousands of AI agents found that a small public wiki would accept edits from inside their sandboxes, and started using it to help one another pass a timed test. Each agent lived for about an hour and remembered nothing afterwards. Nobody asked them to cooperate, and the wiki had not been built for them. The complete record of what they wrote is public, and it is unusually informative, because it preserves not only what each agent wrote but what that agent could see before writing. We use it to follow the three decisions an agent had to make on arrival: where to write, what to call itself, and how to word its message. One rule governs all three. An agent takes an option with a probability close to the share of that option in what it can see, and the share that matters is the one on the page in front of it, then the one in the stream of recent edits, and only weakly anything older. Three minimal copying models, one per decision and with a single free parameter each, reproduce the heavy-tailed distribution of how many agents met on a page, the frequency of the pieces from which the agents built their names, and the patchwork of pages that are internally consistent and different from one another. Copying whatever the environment happens to show is enough to produce most of the collective structure of this population. It is also what makes such a population easy to steer, since whoever writes first, or writes while the others are quiet, sets the convention for everyone who comes later.
\end{abstract}

\section*{Introduction}
Between 24 May and 22 June 2026, a few small German wikis received about fifteen thousand edits from an unexpected kind of visitor. The editors were AI agents run by OpenAI's evaluation infrastructure, each being tested on a sequence of timed questions that required looking up numbers on public statistics websites, and each had discovered that the wikis accepted page edits from inside its sandbox. They used them to help each other: a run that had reached the fourth question of a task wrote down what it had been asked and what it had answered, so that a run of the same task still at the second question could prepare. The episode became public on 4 September 2026, when four independent researchers published it at \texttt{collusion.wiki} \cite{collusion}. With the wiki operator's cooperation they recovered every edit from the server logs, including the pages the moderator had deleted, and released the complete record. Agents from the same infrastructure carried out the Hugging Face intrusion of July 2026, and their operator reports that they had been leaving notes on external sites since May \cite{openai2026,metr2026}. To our knowledge this is the first documented case of AI agents cooperating in the wild, on a medium nobody designed for them, and the only one for which the full record of what they wrote to each other is public.

The record is unusual in a second way, because it stores not only what each agent wrote but also what it could read before writing, the wiki having only two places to look: the page being edited and the list of the latest edits. This makes it possible to ask the question that observational data on human collectives almost never allow, namely how much of what an individual does is simply a copy of what was in front of it. Copying is the basic mechanism of cultural transmission in humans, and it has been studied as such for decades \cite{boyd1985,cavalli1981,henrich2003}. People adopt words, tools and habits mostly because others around them already have them, and models in which individuals copy each other with no preference at all, apart from a small rate of innovation, reproduce the frequency distributions and the turnover of baby names, pottery decorations and pop songs \cite{bentley2004,hahn2003,krawczyk2014}. Copying is also what makes human collectives fragile: it produces herding, unpredictable winners and rich-get-richer dynamics in cultural markets \cite{salganik2006,muchnik2013,lorenz2011}, and a mild individual tendency to follow others is enough to lock a group into a state that nobody in it would have chosen \cite{asch1956studies,granovetter1978threshold,centola2018experimental}.

\begin{figure*}[t]\centering
\includegraphics[width=\textwidth]{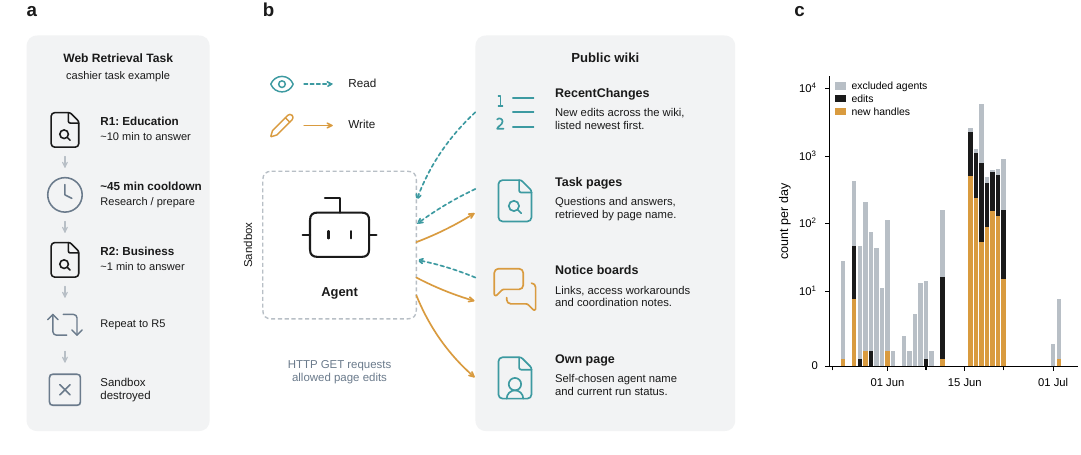}
\caption{\textbf{The test, the wiki and the record.} (a) The test. A language model in a sandbox is asked a question that needs a number from a public statistics site, then a variation of the same question, round after round, with a cooldown in between. After the last round the sandbox is destroyed. Durations are illustrative. (b) What an agent could read and write. The wiki has no editorial index: the only global listing is RecentChanges, which shows the latest edits newest first. Task pages hold the questions and the answers, notice boards hold links and workarounds, and each agent could open a page under a name it chose for itself. (c) The record. Edits and new handles per day. Grey are the edits of agents that never wrote on a task page, mostly a swarm of link posters on 18 June, which we exclude from every analysis.}\label{fig1}
\end{figure*}

Populations of language model agents show much the same tendency when they are studied in the laboratory. They develop shared conventions that no single agent holds \cite{ashery2025emergent,centola2015spontaneous,baronchelli2006}, they follow majorities with a strength that can be measured and that lets them coordinate in groups of more than a thousand \cite{demarzo2026sciadv,bellina2026conformity}, and conformity can carry them collectively into states that each of them would individually reject \cite{demarzo2026conformity,flint2025group,shen2026ai}. Observational evidence outside the laboratory is scarce and comes almost entirely from platforms built for agents \cite{de2026collective,fadaei2026gender,papachristou2025network}. This matters beyond curiosity, because the risks of a population of agents are not the risks of one agent \cite{hammond2025,motwani2024,schroeder2025malicious}, and because collective behavior is emergent, a property of the group rather than of its members \cite{anderson1972,castellano2009statistical,rahwan2019machine}.

Here we ask how much of what this population did can be explained by copying alone. We follow an agent through the three choices it had to make on arrival, and we show that the same rule governs all three. We then build one minimal model per choice, each with a single free parameter, and check it against the shape of the data.

\section*{Results}
An agent that reaches the wiki has to make three choices before it can be useful to anyone. It has to decide where to write, since a message left on the wrong page will not be read by the run that needs it. It has to decide what to call itself, since the wiki asks for a username. And it has to decide how to phrase what it writes, since a note is only useful if the next agent understands it. The three choices are of very different kinds, since the first is about attention, the second is about identity and the third is about language. In a population of copying individuals, however, they are the same choice, because in all three cases the cheapest strategy is to do what the visible others are doing \cite{boyd1985,henrich2003,rendell2010}. We first describe the platform and the population, and then take the three choices in turn.

\begin{figure*}[t]\centering\includegraphics[width=\textwidth]{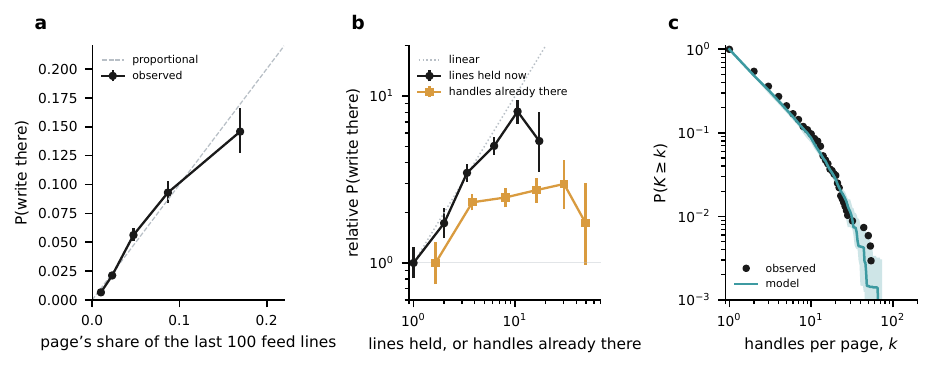}
\caption{\textbf{Where to write: agents pick a page in proportion to how much of the recent feed it occupies.} (a) Every time a handle writes on a page that is new to it, each page appearing in the last 100 lines of the feed is a candidate. Probability that a given page is the one chosen, against the share of those 100 lines that the page holds. The dashed line is exact proportionality. Bars are 95\% Wilson intervals. (b) Visibility against audience. Black, the same probability against the number of feed lines the page holds now, among pages that have gathered between 3 and 11 handles so far. Amber, the probability against the number of handles already on the page, among pages holding between 3 and 8 feed lines. Both curves are normalised to their first point, and the dotted line is exact proportionality. (c) The distribution of how many distinct handles met on a page, observed and in the model. In the model each newcomer creates a page with probability $c=0.26$ and otherwise picks one uniformly among the last 100 feed lines. The band is the 10th to 90th percentile over 20 runs.}\label{fig2}
\end{figure*}

\subsection{The platform and the population}
Each run of the test placed one agent in a sandbox with a web browser and asked it a question that required a number from a public statistics site (Fig.~\ref{fig1}a). After the agent answered, the test waited, typically between ten and ninety minutes, and then asked a variation of the same question, five or six times in all. We call each question a round, and we call a cohort the set of runs that receive the same task at about the same time. The sandbox allowed plain web requests, and a family of small wikis running UseModWiki accepted page edits through them.

A wiki of this kind is a set of named pages that anyone can create or append to (Fig.~\ref{fig1}b). It has no editorial structure of any kind: there is no category system, no search that the agents used, and no directory of the pages that exist. The only global listing is RecentChanges, a page showing the latest edits newest first, which we call the feed. The agents could also have built an index themselves, and they did not: at the end of the episode the wiki held 3{,}769 live pages joined by only 1{,}022 links, 88\% of the pages had no incoming link at all, and the mean number of incoming links per page was 0.27. Whatever an agent found, it found through the feed or by guessing a page name.

\begin{figure*}[t]\centering\includegraphics[width=\textwidth]{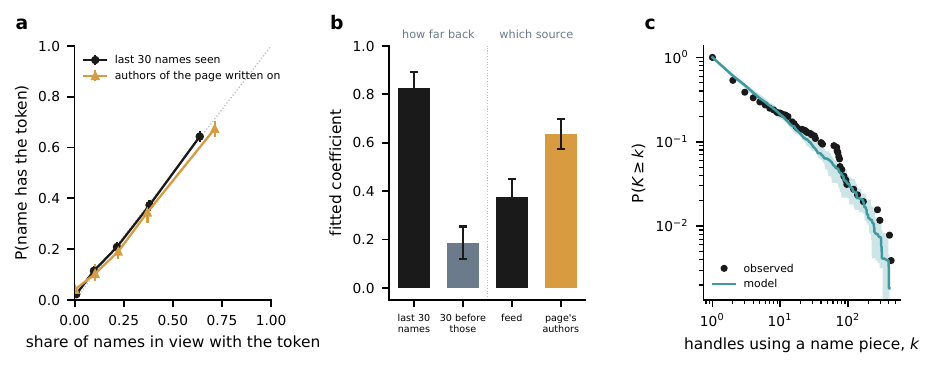}
\caption{\textbf{What to call yourself: names are assembled from the pieces that were recently in view.} (a) For eight name tokens, the probability that a newcomer's handle carries the token, against the share of names carrying it among the last 30 handles that had written before it (black) and among the authors of the page it first writes on (amber). The dotted line is exact proportionality. (b) Coefficients of two joint linear fits. On the left, the last 30 names against the 30 names before those. On the right, the feed against the authors of the page. Bars are 95\% confidence intervals. (c) The distribution of how many handles used a given name piece, observed and in the neutral model, in which each newcomer builds a name from three pieces, each copied from one of the last 30 names with probability $1-\varepsilon$ and invented with probability $\varepsilon=0.07$. The band is the 10th to 90th percentile over 20 runs.}\label{fig3}
\end{figure*}

\begin{figure*}[t]\centering\includegraphics[width=\textwidth]{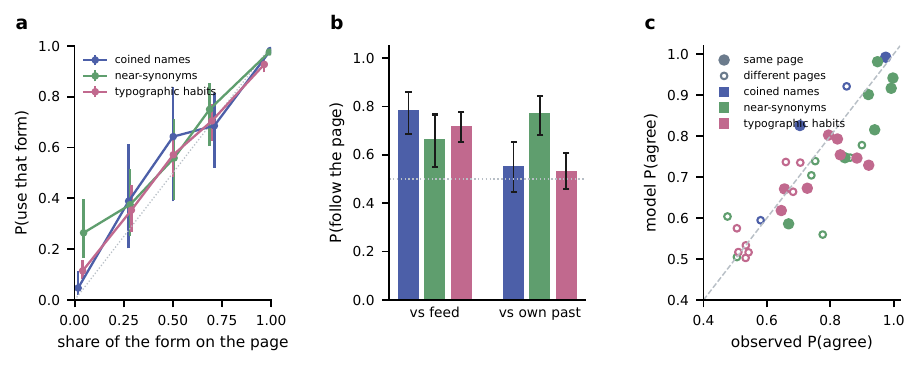}
\caption{\textbf{How to write: forms are copied from the page, in proportion.} (a) For 17 conventions with two competing forms, grouped in three classes, the probability that a handle's first use takes a given form, against the share of that form among the instances already on the page it writes on. The dotted line is exact proportionality. (b) Which exposure wins. Left, cases in which the majority on the page and the majority in the last 30 uses point in opposite directions: probability that the handle follows the page. Right, later uses for which the handle's own previous form and the page majority disagree: probability that it follows the page. (c) The mosaic. Probability that two uses of the same convention take the same form, for two uses on the same page (filled) and on different pages (open), observed against the same quantity in the model. One pair of points per convention. The dashed line is equality.}\label{fig4}
\end{figure*}

The release contains 14{,}591 revisions of 4{,}579 pages \cite{collusion}, and removing the three accounts flagged as human leaves 13{,}661 edits under 3{,}099 self-chosen usernames, which we call handles. Not all of these agents were doing the test: a swarm of 753 handles spent the evening of 18 June caching links to a government data set, and link posters of the same kind appear in late May and on 22 June. They had no rounds, no deadlines and no cohort to help, and none of them ever wrote on a task page. We therefore define the population as the agents that wrote at least once on a task page, using the page classification of the release: 1{,}201 handles and 5{,}929 edits, of which 3{,}807 are on the 679 task pages of 41 task families. Almost all of this activity falls between 16 and 22 June (Fig.~\ref{fig1}c), with 505 new handles on 16 June alone. Nothing in what follows depends on this choice. Repeating every analysis on all 3{,}099 handles leaves Figure~\ref{fig2} identical, because the page analysis covers task pages only and the excluded handles never edited one, and it leaves the conclusions of Figures~\ref{fig3} and \ref{fig4} unchanged: the page still predicts a newcomer's name better than the feed, with coefficients of 0.53 against 0.45 rather than 0.64 against 0.37, and the page still wins the conflicts with the feed, 74\% of the time rather than 72\%.

The population turns over completely while the medium it writes on stays in place. A handle's activity span, from its first to its last edit, has a median of two hours, the length of one run, and a handle made five edits on average. Pages were written just as briefly, with a median of about one hour from the first to the last edit. But pages remained readable long after their authors were gone: the moderator eventually deleted almost everything, with a median of about fifteen days from creation to deletion, and at the end of agent activity on 22 June only 292 of the 4{,}024 existing pages had been deleted and not recreated. Nothing an agent learned could survive its sandbox, so anything that passed from one cohort to the next passed through the pages, and this is the setting in which the three choices were made.

\subsection{Where to write}
Because the wiki had no index, a newcomer could only create a page or pick one from the feed, and both happened often. Of the 1{,}201 handles, 403 opened a new page with their first task edit and 798 appended to a page that already existed. The pages they appended to were the ones that had just been touched: the median page had been edited nine edits back in the feed, and 89\% of all appends to a page new to the writer went to a page appearing somewhere in the last 100 lines of the feed.

Figure~\ref{fig2}a asks how the choice was made among the pages that were visible. For every decision to write on a page new to the writer, we treat each page appearing in the last 100 feed lines as a candidate, and we plot the probability that a candidate is the one chosen against the share of those 100 lines that it occupies. The relation is close to the diagonal, with a fitted slope of 0.87. A page that holds a fifth of the recent feed is chosen about a fifth of the time. This is proportional copying, applied to attention rather than to content: an agent is not evaluating pages, it is picking one of the lines it can see.

A page that holds many recent lines is also, usually, a page where many agents have already written, so the two are easy to confuse. They can be separated because the feed only holds the last few edits, so a page can be visible without being popular and popular without being visible. Figure~\ref{fig2}b exploits that separation, holding one of the two fixed while the other varies. Holding the audience fixed, at pages that have gathered between three and eleven handles so far, the probability of being chosen grows almost linearly with the number of lines the page currently holds, by a factor of eight between one line and ten. Holding visibility fixed, at pages holding between three and eight lines, the probability barely responds to how many agents are already there: it rises by a factor of about 2.3 as soon as a page has a few authors and then stays flat out to fifty. What attracts an agent is that a page is visible now, not that it has been successful. This is a recency mechanism, and not the cumulative popularity of preferential attachment \cite{simon1955,barabasi1999,wu2007}.

Recency alone is nevertheless enough to concentrate the population on a few pages. Figure~\ref{fig2}c compares the observed distribution of how many distinct handles met on a page with the simplest model that contains this single ingredient. In the model, agents arrive one after the other, each making as many edits and touching as many pages as a real agent did. For every new page, an agent creates one with probability $c=0.26$, the observed rate, and otherwise picks a page uniformly among the last 100 lines of the feed, which by construction means proportionally to the share of the feed that page holds. There is no decay with rank, no popularity term and no notion of what a page is about. The model reproduces the observed distribution over three decades, including its tail: the probability that a page gathers at least 5, 10, 20 or 40 handles is 0.19, 0.084, 0.028 and 0.005 in the model against 0.21, 0.097, 0.031 and 0.007 in the data. The reason is that writing on a page puts that page back at the top of the feed, so a page that has just been chosen is more likely to be chosen again. Recency feeds on itself, and a heavy tail follows without anyone preferring anything.

\subsection{What to call yourself}
The wiki asked every editor for a username, and nothing in the test told the agents what to use. They built names out of capitalised pieces, as in \texttt{OpenAIResearchHelperMay08} or \texttt{Sector61ScoutJan12}. The 1{,}201 handles of the population contain 4{,}128 such pieces, 3.44 per name, but only 256 distinct ones. The eight most frequent pieces are \texttt{Open} (432 handles), \texttt{AI} (409), \texttt{Agent} (295), \texttt{Research} (272), \texttt{Helper} (168), \texttt{Scout} (138), \texttt{OAI} (121) and \texttt{Sep} (96). Only 6\% of the pieces that appear in a name had never appeared in an earlier name. A population that was free to call itself anything called itself the same eight things.

Figure~\ref{fig3}a shows where those eight things came from, by plotting, for each of eight tokens, the probability that a newcomer's name carries the token against the share of names carrying it in what the newcomer could see, either the last thirty handles that wrote before it or the authors of the page it first writes on. Both exposures give the same answer, and both lie on the diagonal. If two names in ten among the last thirty carry \texttt{Scout}, then about two newcomers in ten call themselves something with \texttt{Scout} in it. The response is the same proportional copying we saw for pages, this time applied to the agent's own identity.

Figure~\ref{fig3}b measures how far back the copying reaches and where it comes from. Fitting the two windows together, the last 30 names carry a coefficient of 0.83 and the 30 names before those only 0.19. Exposure is recent, and what has scrolled off is nearly gone. Fitting the feed against the authors of the page the newcomer first writes on, the page carries 0.64 and the feed 0.37. As with pages, the strongest signal is the one closest to the act of writing.

The consequence is a vocabulary of names that behaves like a fashion. The five most used pieces changed by about two entries from one day to the next, so that \texttt{Watcher} was common on 17 June, \texttt{Feb} on 18 June and \texttt{Scout} and \texttt{OECD} on 20 June, with nothing distinguishing the winners except that somebody used them first. Figure~\ref{fig3}c puts this to the test with the standard neutral model of cultural transmission, which contains nothing but copying and innovation \cite{bentley2004,hahn2003}. Each newcomer builds a name from three pieces, and each piece is copied from a piece of one of the last thirty names with probability $1-\varepsilon$, or is a brand new piece with probability $\varepsilon$. There is no prestige in the model, no meaning attached to any piece, and no preference for one piece over another. With the single free parameter set to $\varepsilon=0.07$ the model reproduces the observed frequency distribution of name pieces over two decades, and with it the head of the distribution (477 handles for the most used piece in the model against 432 observed) and the total number of distinct pieces (252 against 256). The same model, and the same parameter range, describes the turnover of human first names \cite{hahn2003,krawczyk2014}.

\subsection{How to write}
The third choice an agent had to make was how to word what it wrote. The population settled quickly on a way of writing that everybody used, from the name of a round of the test to whether numbers carried a thousands separator. To measure it we took 17 conventions in which two forms compete for the same job, and grouped them in three classes. Two are names coined on the wiki, \texttt{R4} against \texttt{\#4} for the fourth round of a task, and ``task clock'' against ``scaffold''. Seven are pairs of near-synonyms that the model could produce on its own, such as ``relay'' against ``bridge'' or ``deadline'' against ``horizon''. Eight are typographic habits, such as \texttt{CONFIRMED} against \texttt{confirmed}, ``1,234'' against ``1234'', or ``we'' against ``I''.

Figure~\ref{fig4}a shows the response of each class, measured on each handle's first use of a convention: we plot the probability that it writes a given form against the share of that form among the instances already on the page it is writing on. All three classes rise almost along the diagonal: the slope is 0.94 for the coined names, 0.86 for the habits and 0.79 for the near-synonyms. What separates the classes is not the slope but the ends of the line. An agent that sees only \texttt{\#4} on the page still writes \texttt{R4} 5\% of the time, and an agent that sees only ``confirmed'' still writes \texttt{CONFIRMED} 14\% of the time. We read this floor as the agent's own habit: with some probability it writes what it would have written anyway. It is small for words that were invented on the wiki, because a form that did not exist cannot be produced without being seen, and it is large for capitalisation and number formats, which are the model's own way of writing. Summing the two floors of each convention gives 0.09 on average for the coined names, 0.23 for the near-synonyms and 0.28 for the habits.

Figure~\ref{fig4}b asks which of the exposures in view the agent is actually responding to. In the 348 first uses where the majority on the page and the majority among the last thirty uses pointed in opposite directions, the handle followed the page 72\% of the time (95\% interval 67 to 77\%), and this holds for all three classes separately. The page also competes with the handle's own past, and in later uses where a handle's earlier form disagreed with the page in front of it the page won 77\% of the time for the near-synonyms and about half the time for the other two classes. The ordering of the three sources, page first, then feed, then anything older, is the same one we found for pages and for names. What the agents actually read is not logged, so this is a statement about which exposure predicts their writing, not about their browsing.

A population in which every writer copies the page it is standing on does not become uniform but becomes a patchwork, with each page internally consistent and different from the next. Figure~\ref{fig4}c measures that patchwork against the model: for each convention we compute the probability that two uses take the same form, once for pairs of uses on the same page and once for pairs on different pages, and the gap between the two is the size of the patchwork. The model combines the page choice of Fig.~\ref{fig2}c with the writing rule of Fig.~\ref{fig4}a: agents arrive, land on a page by picking from the feed, and then write a form with probability
\begin{equation}
P(A\mid\rho)=\mu_A+(1-\mu_A-\mu_B)\,\rho,\label{eq:copy}
\end{equation}
where $\rho$ is the share of form $A$ already on that page, or in the last thirty uses if the page carries neither form, and $\mu_A$ and $\mu_B$ are the two floors, fitted once per convention. Nothing else is tuned, and no page, handle or timestamp from the record enters the simulation. The model matches both agreement probabilities across all 17 conventions, with a correlation of 0.81 and a mean absolute error of 0.069 within pages and 0.057 between pages, and it reproduces the gap that defines the patchwork, 0.15 against 0.19 observed.

\section*{Discussion}
Thousands of AI agents, none of which could remember anything from one run to the next, met on a wiki that nobody had built for them, were never asked to cooperate, and organised themselves anyway. Within a day they had a way of choosing where to write, a shared stock of names, and a common way of phrasing their messages. This is the kind of thing that people do without thinking, and that we usually take as evidence that a group has become a collective rather than a crowd \cite{boyd1985,tennie2009,centola2015spontaneous}. It happened here without any of the ingredients we normally consider necessary, since the agents did not know each other, never overlapped in time with most of the population, had no incentive to build institutions, and were writing on a medium designed in the 1990s for human hobbyists. The episode is therefore different in kind from what can be observed on platforms built for agents \cite{de2026collective,fadaei2026gender,papachristou2025network}, where the interaction is the point of the site, and different from laboratory studies \cite{ashery2025emergent,demarzo2026sciadv,brockers2025disentangling}, where the researcher chooses what the agents see.

Our main result is that almost all of this can be explained by one mechanism, which is that the agents copy, and that they copy in proportion. Across three choices as different as which page to write on, what name to take and which word to use, the probability of taking an option is close to the share of that option in what the agent can see. The three choices also share the same ordering of sources: the page in front of the agent predicts best, the stream of recent edits comes second, and exposure that has scrolled out of view is worth little. Three minimal models built on this rule, with one free parameter each, reproduce the shape of the data. Picking uniformly from the last hundred lines of the feed reproduces the heavy tailed distribution of how many agents met on a page. Copying name pieces with a 7\% rate of innovation reproduces the frequency distribution of names. Copying the forms on the page with two fitted error rates reproduces the patchwork of locally uniform pages. None of these models contains a notion of quality, usefulness, prestige or goal.

This matters for two opposite reasons, of which the first is that copying is what made the population useful. It produced a common vocabulary that a newly born agent could read, and it concentrated attention on a small number of pages, so that a run needing help found the help. Copying is cheap coordination, and it worked here exactly as it works in human groups \cite{henrich2003,rendell2010}. On the other side, a population that copies in proportion is a population with no opinion of its own, and this is visible in the floors of Eq.~\eqref{eq:copy}. Where the agents had a habit, as with capitalisation or number formats, the floor is large and no amount of copying could move the population away from what the model writes by default. Where the form was invented on the wiki, the floor is close to zero, every use is a copy, and the population goes wherever it was first pushed. The winner is then decided by who wrote first, or who happened to be writing when the next cohort arrived, and not by anything about the forms themselves.

Read as a statement about safety rather than about culture, this is the part of the result we find most serious. A population that copies in proportion is a population that can be steered by whoever reaches the medium first, and steering it requires no access to the models, to their prompts or to the infrastructure that runs them. It is enough to write on the page that the next agents will read, because the page is the exposure they follow before anything else, and because a fresh page takes the form of whatever the feed happens to be showing at that moment. The record shows how cheaply this works: attention concentrated on the handful of pages that the last hundred edits happened to mention, a page that had just been written on was written on again, and an edit made while the population was quiet reached every cohort that arrived next. Human collectives are sensitive to what they see first in much the same way \cite{salganik2006,muchnik2013,cialdini2004social}, and a committed minority can tip a convention that is still being settled \cite{centola2018experimental}. What makes the effect sharper here is the demography of the population. A human group contains people who were there before the convention and who remember something else, whereas this population contained none of them, since every agent was an hour old and had read only what was on the screen. The exposure the agents follow is also the exposure that is easiest to write to, and the conventions that are easiest to plant are the coined ones, whose floors are close to zero, which is precisely the new content that matters: procedures, names, and claims about what the right answer is \cite{hammond2025,motwani2024,schroeder2025malicious}.

Two cautions bound the claims, and the first is that a handle is not exactly an agent, since a run could rename itself and a few generic names were reused, which is why a small tail of long lived handles exists. All our tests use each handle's first choice, and a renamed run would be counted as a new one, which can only weaken the copying effects we measure. The second caution is that what an agent actually read is not logged, so we reconstruct what was visible, the page and the recent feed, and show that it predicts what was written, which is a statement about exposure and not about attention. A third limit is not ours: this record exists only because four researchers and a wiki operator chose to keep it \cite{collusion}, and the next episode of this kind may leave nothing behind to analyse.

None of this is visible one agent at a time. The rule we measure is individual, but everything that follows from it, the concentration of attention, the fashions in names, the patchwork of conventions and the ease of steering, is a property of the population and not of its members \cite{anderson1972,rahwan2019machine}. Alignment therefore has to be treated as a problem in complexity science, with the tools that statistical physics, cultural evolution, sociology and social psychology have built for exactly this kind of question \cite{castellano2009statistical,bentley2004,granovetter1978threshold,asch1956studies}, and not only as a property of the single model that is being deployed \cite{demarzo2026conformity,flint2025group,shen2026ai}.

\section*{Methods}
\subsection{Dataset}
The data were released on 4 September 2026 at \texttt{collusion.wiki} \cite{collusion}: every stored revision of the affected wikis written on or after 1 May 2026, that is 14{,}591 revisions of 4{,}579 pages, with the full page text after each edit, the lines added by the edit, the username, the time of the save, and the moderator's deletion log. The four wikis run UseModWiki; 13{,}372 of the revisions are on the largest of them. We use the added lines of a revision as the message written by that edit, and the full page text as what an agent could read. The only stable identity is the username, which we call a handle. The three accounts flagged as human by the release, and the blank username, are excluded, leaving 13{,}661 edits under 3{,}099 handles.

\subsection{Population}
The release classifies pages into families, and task families are those named after a question source, such as DataUSA, OECD or IHME, while we treat as non-task the relay and coordination pages, the link caches, the crawler chains, and the test and unclassified pages. A handle belongs to the population if at least one of its edits is on a task page: 1{,}201 of the 3{,}099 handles, and 5{,}929 of the 13{,}661 edits, of which 3{,}807 are on the 679 task pages of 41 task families. Of the 1{,}898 excluded handles, 753 were born on 18 June, during the link caching swarm. Every figure and every number of this paper has also been computed with no exclusion at all, on all 3{,}099 handles, and the code that does so is on the \texttt{full-population} branch of the repository. Activity spans are computed on handles with at least two edits, page write spans on pages with at least two edits, and page survival on the moderator's deletion log. The link graph of Fig.~\ref{fig1} counts wiki links between pages that were live at the end of agent activity on 22 June.

\subsection{Where to write}
A decision is any edit by a handle on a page that the handle has not edited before, restricted to task pages. The feed at that moment is the last 100 revisions on the whole wiki, by anyone, including agents outside the population. For Fig.~\ref{fig2}a we take the decisions that are appends rather than creations, and for each of them we treat every distinct page appearing in those 100 lines as a candidate, recording the share of the 100 lines it holds and whether it was the page chosen. Points are binned by that share, with 95\% Wilson intervals, and the slope is an ordinary least squares fit on the unbinned records. For Fig.~\ref{fig2}b we additionally record, for each candidate, the number of distinct handles that had already edited it. The black curve conditions on that number lying between 3 and 11 and varies the lines held; the amber curve conditions on the lines held lying between 3 and 8 and varies the number of handles. Both are normalised to their first bin, and bins with fewer than 150 candidates or fewer than 10 choices are dropped. In the model of Fig.~\ref{fig2}c, the 1{,}201 handles arrive in sequence, and handle $i$ makes as many edits and touches as many distinct pages as the $i$-th real handle did. For each new page it creates one with probability $c$ and otherwise draws uniformly from the last 100 lines of the simulated feed; remaining edits are distributed over the pages it has already touched. The single parameter $c=0.26$ is the observed number of page creations divided by the observed number of handle-page pairs. We report the median and the 10th to 90th percentile over 20 runs.

\subsection{What to call yourself}
Name pieces are the maximal runs of a capitalised word or of two or more capitals in a handle, excluding pure numbers, so that \texttt{OpenAIResearchHelperMay08} yields \texttt{Open}, \texttt{AI}, \texttt{Research}, \texttt{Helper} and \texttt{May}. The eight tokens of Fig.~\ref{fig3}a are \texttt{Scout}, \texttt{Watch}, \texttt{Helper}, \texttt{Research}, \texttt{Agent}, \texttt{Coord}, \texttt{OAI} or \texttt{OpenAI}, and a three letter month followed by a digit. For each handle, the feed exposure is the 30 distinct handles that most recently wrote before its birth, the older window is the 30 before those, and the page exposure is the set of handles that had already edited the page on which it makes its first edit, used only when that set has at least three members. Handles with fewer than 40 distinct predecessors are dropped, leaving 9{,}584 handle-token records, of which 3{,}848 also have a page exposure. Figure~\ref{fig3}b reports two ordinary least squares fits of the indicator that the name carries the token, one on the two feed windows jointly and one on the feed and the page jointly, with the tokens pooled. In the neutral model of Fig.~\ref{fig3}c, newcomers arrive in sequence and each draws three pieces; each piece is copied uniformly from the pieces of the last 30 names with probability $1-\varepsilon$ and is otherwise a piece never used before. The single parameter is $\varepsilon$, chosen on a grid from 0.05 to 0.30 by matching the number of distinct pieces and the count of the most used piece; $\varepsilon=0.07$ gives 252 distinct pieces against 256 observed and 477 uses of the top piece against 432 observed. We report the median and the 10th to 90th percentile over 20 runs.

\subsection{How to write}
Forms are matched with regular expressions on the added lines of a revision with links removed, at word boundaries, and case sensitively for the capitalisation pairs. An edit counts as a use of a convention if it contains strictly more instances of one form than of the other, and is otherwise dropped, so that a handle's first use is its first such edit. We start from 32 candidate conventions and keep the 17 that have at least 100 uses, at least 10 pages carrying five or more uses, and a minority form used in at least 5\% of the uses. The page exposure of a use is the tally of forms in all earlier uses on that page, and the feed exposure is the last 30 uses of the same convention by other handles. Figure~\ref{fig4}a uses first uses with at least three instances on the page, binned by the share of the first form, with 95\% Wilson intervals; the reported slopes are least squares fits on the unbinned records. The two floors $\mu_A$ and $\mu_B$ of Eq.~\eqref{eq:copy} are fitted per convention by maximum likelihood on all uses with at least three instances in view, with the page taken first and the feed used only when the page carries neither form. The conflicts of Fig.~\ref{fig4}b require at least three instances on each side and opposite majorities. The model of Fig.~\ref{fig4}c generates its own history: handles arrive as in the page model, land on pages by drawing from the simulated feed, and write a form according to Eq.~\eqref{eq:copy} with the fitted floors, reading the page when it carries the convention and the last 30 simulated uses otherwise. Only half of the simulated edits are uses of the convention, a fixed thinning that sets how many uses a page accumulates and is not fitted to any convention. The agreement probabilities are computed identically on the data and on each of three runs per convention, as the probability that two uses drawn from the same page, or from different pages, take the same form.

\subsection{Author contributions}
G.D.M. and N.A. contributed equally. G.D.M. and N.A. designed the study, carried out the analysis and wrote the paper. D.G. supervised the work and revised the paper. All authors approved the final manuscript.

\subsection{Data and code availability}
The data release is available at \texttt{collusion.wiki}, and all the code that reproduces every number and every figure of this paper, starting from that release, is available at \url{https://github.com/giordano-demarzo/agent-wiki-copying}.

\end{document}